# Carrier Injection and Scattering in Atomically Thin Chalcogenides

Song-Lin Li[1,2,*] and Kazuhito Tsukagoshi[1*]

*[1]WPI Center for Materials Nanoarchitechtonics (WPI-MANA) and [2]International Center for Young Scientists (ICYS), National Institute for Materials Science, Tsukuba, Ibaraki 305-0044, Japan*
Email: songlinli@gmail.com; tsukagoshi.kazuhito@nims.go.jp

**Abstract:** Atomically thin two-dimensional chalcogenides such as $MoS_2$ monolayers are structurally ideal channel materials for the ultimate atomic electronics. However, a heavy thickness dependence of electrical performance is shown in these ultrathin materials, and the device performance normally degrades while exhibiting a low carrier mobility as compared with corresponding bulks, constituting a main hurdle for application in electronics. In this brief review, we summarize our recent work on electrode/channel contacts and carrier scattering mechanisms to address the origins of this adverse thickness dependence. Extrinsically, the Schottky barrier height increases at the electrode/channel contact area in thin channels owing to bandgap expansion caused by quantum confinement, which hinders carrier injection and degrades device performance. Intrinsically, thin channels tend to suffer from intensified Coulomb impurity scattering, resulting from the reduced interaction distance between interfacial impurities and channel carriers. Both factors are responsible for the adverse dependence of carrier mobility on channel thickness in two-dimensional semiconductors.

## 1. Introduction

As microelectronics approaches its physical limit,[1-3] two-dimensional (2D) chalcogenides have emerged as a promising candidate for the ultimate atomic field-effect transistor (FET) technology after silicon (Si).[4-6] The combination of four features in the 2D chalcogenides lays the foundation for technologically viable atomic electronics. First, the intrinsic semiconducting nature, contrasting graphene's metallicity,[7-14] enables high on/off current ratios, a basic requirement for logic operation. Second, the planar 2D structure offers compatibility to optical lithography and large-scale fabrication, rivaling 1D nanostructures such as carbon nanotubes. Moreover, the ultrasmall thickness allows for more aggressive device downscaling than silicon and thus higher density in integration. Fourth, the atomic flatness makes them immune to surface-roughness-induced carrier scattering so that they can overcome the limitation of channel thickness confronted by Si FETs.[15-17]

Extensive research attention has been devoted to 2D chalcogenides spanning from synthesis[18-24] and characterization[25,26] to electrical,[27-33] optoelectronic[34-36] and photovoltaic[37-41] applications, and further to novel valley physics[42-48] and nonlinear optics.[49,50] At present, reviews of various topics are available.[51-55] In this brief review, we focus on their application as FET channels and, more specifically, on the factors resulting in the adverse thickness dependence of the electrical performance of atomically thin 2D $MoS_2$ flakes.

## 2. Dependence of Performance on Thickness

High-quality atomic chalcogenide crystals were first isolated by Novoselov et al. in 2005,[5] immediately after the isolation of graphene. In the beginning, they did not generate broad interest, likely owing to their considerably low carrier mobility ($\mu$) of <10 $cm^2V^{-1}s^{-1}$, as compared with graphene. They attracted renewed attention after a series of reports claiming excitingly high carrier mobility in top high-k dielectric gated monolayer $MoS_2$ FETs.[56,57] Although it was pointed out that these results were overestimated,[58] much research effort was continuously exerted because it was realized that the intrinsic semiconducting nature of 2D chalcogenides can complement the metallic behavior of graphene.

Figure 1(a) shows the optical image and device structure of $SiO_2$-supported 2D $MoS_2$ FETs. Reproducible data show that the carrier mobility of $MoS_2$ monolayers normally falls in the range of 1–10 $cm^2V^{-1}s^{-1}$. The ultrathin 2D $MoS_2$ flakes exhibit markedly degraded electrical performance when compared with corresponding bulks with a high mobility of ~200 $cm^2V^{-1}s^{-1}$.[59] More extensive experiments show that the carrier mobility shows a heavy dependence on channel thickness.[60,61] Figure 1(b) shows a summary of the evolution of mobility for $MoS_2$ FETs with channel thickness. In the FET measurement geometry, the mobility exhibits two varying trends separated by a critical thickness of about 14 layers. Above the critical thickness, mobility increases monotonically with decreasing channel thickness. Such an increasing trend of mobility results from the reduction in c-axis access resistance,[62] which is nontrivial. Below the critical thickness, however, the mobility decreases from ~180 to ~20 $cm^2V^{-1}s^{-1}$ with further decreasing channel thickness. In such an ultrasmall thickness regime, the varying rate of mobility of 2D $MoS_2$ is relatively weaker than that of silicon, in which an extremely strong power-law ($\mu \sim t^{-6}$) dependence on channel thickness ($t$) dominates as $t < 4$ nm owing to the carrier scattering from surface roughness (SR).[15-17] Given the atomically sharp surfaces shown by the 2D MoS$_2$, the SR scattering mechanism is obviously not the origin of the mobility degradation observed in the ultrathin 2D chalcogenides. The intriguing thickness dependence and performance degradation remain to be elucidated.

To resolve the above issue, we have performed a combined experimental and theoretical study of atomically thin MoS$_2$ FETs with varying channel thickness.[61,63] We revealed that the electrode/channel contacts and carrier scattering mechanisms are responsible for the adverse thickness dependence. On the one hand, the electrode/channel contact is an extrinsic factor affecting device performance in the way that the Schottky barrier height at the contact area is increased in thin channels owing to the expansion of the channel bandgap caused by quantum confinement,[64] which hinders carrier injection and degrades device current. On the other hand, the scattering from Coulomb/charged impurities (CIs) is an intrinsic factor because thin channels tend to experience intensified Coulomb impurity scattering, resulting from the reduced interaction distance between interfacial CIs and channel carriers. Both factors are responsible for the adverse dependence of mobility on channel thickness for 2D semiconductors.

## 2.1 Electrode/semiconductor contact

Since FETs operate with two metal electrodes (source and drain), the electrode/channel contacts play an important role in device performance. To understand the interfacial electrical properties between metallic electrodes and 2D chalcogenides, we perform a systematic thickness scaling study of Au/2D MoS$_2$ interfaces. Transfer line measurement [Fig. 1(a)] is used to extract the contact resistance ($R_c$, in unit of $\Omega$ cm) together with the intrinsic channel resistance ($R_s$, in unit of $\Omega$/square). Meanwhile, a top-Au-contacted, bottom-SiO$_2$-gated FET structure is employed to determine the gate dependences of $R_c$ and $R_s$. The linear drain current ($I_d$) *vs* drain voltage ($V_d$) is observed in thermally annealed samples [Fig. 2(a)], indicating excellent contact between Au and MoS$_2$. Figure 2(b) shows a typical transfer line plot for a sample under different gating conditions. The electrical parameters $R_c$ and $R_s$ are extracted from the intercepts and slopes of the linear fittings.

Figure 2(c) shows a summary of $R_c$ *vs* gate bias ($V_g$-$V_t$) for MoS$_2$ thicknesses from 5 to 1 layer. Two features are shown in the $R_c$ curves. First, $R_c$ highly depends on gate bias and is largely reduced by 1–2 orders in magnitude depending on sample thickness. Second, thinner

MoS$_2$ flakes result in higher $R_c$ values, indicating similar physical characteristics at the interface of 2D semiconductors to 1D carbon nanotubes,[65] *i.e.*, dimension reduction leads to barrier enhancement.[65] Figure 2(d) shows the intrinsic values of $R_s$ extracted from the transfer line measurement where $R_c$ is deducted. As expected, $R_s$ decreases with increasing gate bias, as a consequence of field-effect gating. The inset of Fig. 2(d) shows a plot of the intrinsic carrier mobility as a function of thickness, indicating that the mobility dependence is an inherent behavior regardless of contact condition.

As far as electrical engineering is concerned, the specific contact resistivity ( $\rho_c$, in unit of $\Omega$ cm$^2$) is more explicit for characterizing contact quality because it excludes the effect of current crowding. The relationship of $\rho_c$ to $R_c$ and $R_s$ can be derived from a resistor network model[66] where the electrode/channel stack is divided into infinite impedance elements [lateral resistor d$R$ and vertical conductor d$G$, Fig. 3(a)]. The impedance elements for the channel and interface are given by $dR = R_s w^{-1} dx$ and $dG = \rho_c^{-1} w dx$, where $w$ and $x$ are the channel width and coordinate, respectively. The metal electrodes are treated as ideal conductors with $R=0$. The lateral channel current is defined as $i(x)$. At the electrode/channel interface, the vertical potential and current density are defined as $u(x)$ and $j(x)$. According to Kirchhoff Circuit Laws, $u(x)$ and $i(x)$ in the stack can be described by the differentiate equations

$$\begin{cases} u(x+dx)-u(x)=i(x)\cdot dR \Rightarrow du/dx = iR_s/w \\ i(x+dx)-i(x)=u(x)\cdot dG \Rightarrow di/dx = uw/\rho_c \end{cases} , \qquad (1)$$

which lead to the reciprocal relations between $u(x)$ and $i(x)$,

$$\begin{cases} \dfrac{d^2 i}{dx^2} = \dfrac{R_s}{\rho_c} u \\ \dfrac{d^2 u}{dx^2} = \dfrac{R_s}{\rho_c} i \end{cases} . \qquad (2)$$

Using the boundary conditions $i(0)=0$, $u(0)\neq 0$, the relationship among $\rho_c$, $R_c$, and $R_s$ is derived as

$$R_c w = \sqrt{R_s \rho_c} \coth(l_c \sqrt{R_s/\rho_c}) . \qquad (3)$$

To illustrate the current crowding effect, Figs. 3(b) and 3(c) show the current and potential distributions along the contact length by setting $l_c \sqrt{R_s/\rho_c} = 1$. Apparently, the current injection is not uniform along the electrodes.

Figure 4(a) shows the extracted $\rho_c$ values under different gate conditions for the thickness-varied samples. A remarkable finding emerges when plotting $\rho_c$ *vs* $MoS_2$ thickness [Fig. 4(b)]. Two opposite $\rho_c$ trends appear in different thickness regimes, with a positive slope in the 3D regime and a negative slope in the 2D regime, resulting in a dip around 5 layers. Detailed analysis indicates that the negative $\rho_c$ slope in the 2D regime results from the variation in interfacial barrier height owing to bandgap expansion in the quantum confinement regime.[63] In contrast, the positive $\rho_c$ slope in the 3D regime originates from the reduction in the number of upper inactive layers as channel thickness decreases [Fig. 4(c)], a characteristic of the anisotropic transport behavior.

According to the theories on carrier injection at the metal/semiconductor interfaces,[67,68] the injection current is determined by the barrier width at the interface, which is a parameter that is tunable by adjusting the gate bias (i.e., carrier concentration $n$). Figure 4(d) illustrates the injection mechanisms at different barrier widths. At low gate bias, a thermal emission (TE) process dominates the injection where carriers have to surmount the full height of the barrier. In this case, $\rho_c$ is gate-independent and behaves as[67,68]

$$\rho_c = \frac{k}{A*Te}\exp\left(\frac{e\phi_B}{kT}\right) \quad , \tag{4}$$

where $k$, $T$, $e$, and $\phi_B$ are the Boltzmann constant, temperature, elementary charge, and interfacial Schottky barrier height, respectively; the Richard constant $A* = 8\pi m*ek^2h^{-3}$ with $m*$ the effective mass and $h$ the Planck constant.

At high gate bias, the induced dense carriers considerably decrease the barrier width and increase the tunneling probability of carriers. Then, thermally assisted tunneling (also called thermal field emission, TFE) dominates the injection process. In this case, $\rho_c$ becomes gate-dependent and follows[67,68]

$$\rho_c = \frac{k\sqrt{E_{00}}\cosh(E_{00}/kT)\coth(E_{00}/kT)}{A*Te\sqrt{\pi e(\phi_B - u_f)}}\exp\left[\frac{e(\phi_B - u_f)}{E_{00}\coth(E_{00}/kT)} + \frac{eu_f}{kT}\right] \quad , \tag{5}$$

where $u_f$ is the chemical potential and $E_{00} = eh\sqrt{n_{3D}/(4m*\varepsilon)}$ is a gate-bias-related parameter with $\varepsilon$ being the permittivity. In the first-order approximation, $n_{3D}$ can be calculated using $n_{3D} = V_gC_{ox}/t$ for few-layer samples. As seen in Fig. 4(e), reasonable agreement between the experiment and the calculation is reached.

The above gate bias-$\rho_c$ relationship offers a convenient way of estimating the barrier height $\phi_B$. As evident in Fig. 4(e), all $\rho_c$ data fit well to the TFE injection theory, which enables us to extract the $\phi_B$ values for all the samples. Figure 4(f) illustrates the derived $\phi_B$ *vs* channel bandgap for MoS$_2$ ranging from 1 to 5 layers. A linear fit reveals a slope of 0.46, indicating that the upshift of the conduction band $E_c$ is approximate to the downshift of the valance band $E_v$. In other words, nearly half of the bandgap expansion due to thickness reduction is used to build up the interface barrier. Such behavior resembles that observed in 1D carbon nanotubes,[65] suggesting similar electrostatic equilibrium dynamics between 2D and 1D semiconductors. Figure 4(g) depicts an evolution diagram for energy-level alignment to summarize the thickness scaling effect on the interfacial potential barrier. The barrier height increases from 0.33 to 0.65 eV as the MoS$_2$ thickness decreases from 5 to 1 layer. The increase in barrier height with decreasing channel thickness explains the thickness dependence of contact resistance and constitutes the extrinsic origin of low mobility in FETs with 2D chalcogenide channels, which can be basically suppressed by thermal annealing and/or energy-level matching.[62]

## 2.2 Carrier scattering mechanism

Aside from the extrinsic origin, we next show that CI scattering is the intrinsic origin of the performance degradation in atomically thin FETs. CI scattering has been investigated in silicon FETs,[69] superlattices,[70,71] and graphene[72,73], with specific approximations in each case. No generalized models have been discussed for a common dielectric/channel/dielectric trilayer system [*e.g.*, air/MoS$_2$/SiO$_2$ structure, Fig. 5(a)] with finite channel thickness $t$, asymmetric dielectric surroundings ($\varepsilon_2 \neq \varepsilon_3$), or lopsided carrier distribution. For bulk Si FETs[69] [(Fig. 5(b)] and graphene[72,73] [Fig. 5(c)], approximations of $t = \infty$ and $t = 0$ are adopted, respectively. A more exclusive condition used in the single-atom-thick graphene includes the adoption of a pulselike carrier distribution with a Dirac $\delta$ function.[72,73] For superlattices[70,71] [Fig. 5(d)], a tunable channel thickness is considered but symmetric carrier distributions (trigonometric) and dielectric surroundings ($\varepsilon_2 = \varepsilon_3$) are often employed. Since the configurative conditions strongly modify the screening and polarization of carriers, the specific approximations used in previous models restrict their direct application to the common trilayer structure. Without strictly considering the configurative differences, rigidly applying previous models may cause a large deviation.

We developed a generalized CI scattering model in an effort to cover all configurative conditions. We begin with a lopsided carrier distribution in a finite channel thickness by adopting an envelope electron wavefunction following the form of bulk Si FETs,[69]

$$\phi(z) = \begin{cases} 0, \ |z| > t/2 \\ (b^3/2)^{1/2}(z+t/2)e^{-b(z+t/2)/2}, \ |z| \leq t/2 \end{cases} , \qquad (6)$$

where $z$ is the position in the channel and $b$ is a variational parameter that depends on channel thickness $t$ and gate bias $V_g$. The carrier distribution is expressed as $g(z) = |\phi(z)|^2$. In such a wavefunction, $b$ determines carrier distributions for different $t$ and $V_g$ values. The accurate form of $b$ can be derived from the energy minimum principle and the exact expression would be rather complicated. In our model, a phenomenological relation is introduced by assuming $b = kV_g/t + b_{bulk}$ with $k$ being a tunable coefficient in the unit of V$^{-1}$. Such a form, although simple, can bridge $t$ in the entire channel range and correlate $V_g$, which well describes the dependence of carrier distribution on these two factors. It is easy to justify that $b \rightarrow b_{bulk}(\infty)$ as $t \rightarrow \infty \ (0)$, representing the bulk (or the pulselike $\delta$ function) limit. We find that $k = 1/2$ is an appropriate value and is used in all calculations.

In the next step, the asymmetric dielectric surroundings are strictly considered. To this end, we need to derive the Coulomb force between two point charges in a trilayer system, which is generally solved by a mirror imaging method (Fig. 6). The effect of dielectric asymmetry on CI scattering will be manifested in the dielectric polarization function and scattering matrix elements. To calculate the Coulomb force between two point charges, one charge is set static and the other is constantly mirror-imaged by the two boundaries, as shown in Figs. 6(b) and 6(c). This imaging process produces infinite imaging charges, and the final expression of the Coulomb force is the sum of an infinite series. The expressions of Coulomb forces between two point charges at different locations are listed in Eqs. 7 and 8, where $\bar{r}$ denotes the planar coordinate in the channel plane and $\bar{\rho} = (\bar{r}, z)$ is the spatial coordinate.

Specifically, two cases need to be considered in our model: (1) both point charges are located at the center [Fig. 6(b)], which reflects the carrier-carrier interaction and the polarization of carriers, for deriving the dielectric polarization function $\varepsilon(q,T)$; (2) one point charge is located at the center and the other is on the right [Fig. 6(c)], which reflects the interaction of an external charged impurity with a carrier, for deriving the scattering matrix elements $U_j(q)$. In case 1, the Coulomb force is

$$F_{CC}(\vec{\rho}_a, \vec{\rho}_b) = \sum_{n=-\infty}^{\infty} \frac{e^2 \xi^{|n|}}{\varepsilon_1 [(\vec{r}_a - \vec{r}_b)^2 + (z_a - z_b - 2nt)^2]^{1/2}} + \frac{\varepsilon_1 - \varepsilon_3}{\varepsilon_1 + \varepsilon_3} \sum_{n=0}^{\infty} \frac{e^2 \xi^n}{\varepsilon_1 [(\vec{r}_a - \vec{r}_b)^2 + (z_a + z_b - (2n+1)t)^2]^{1/2}}$$

$$+ \frac{\varepsilon_1 - \varepsilon_3}{\varepsilon_1 + \varepsilon_3} \sum_{n=0}^{\infty} \frac{e^2 \xi^n}{\varepsilon_1 [(\vec{r}_a - \vec{r}_b)^2 + (z_a + z_b + (2n+1)t)^2]^{1/2}} \tag{7}$$

where $\xi = \frac{\varepsilon_1 - \varepsilon_2}{\varepsilon_1 + \varepsilon_2} \frac{\varepsilon_1 - \varepsilon_3}{\varepsilon_1 + \varepsilon_3}$ with $\varepsilon_i$ ($i = 1, 2, 3$) the relative dielectric constants for the $i$th layers. In case 2, the force is

$$F_{LC}(\vec{\rho}_a, \vec{\rho}_b) = \frac{2}{\varepsilon_1 + \varepsilon_2} \sum_{n=0}^{\infty} \frac{e^2 \xi^n}{[(\vec{r}_a - \vec{r}_b)^2 + (z_a - z_b - 2nt)^2]^{1/2}}$$

$$+ \frac{2}{\varepsilon_1 + \varepsilon_2} \frac{\varepsilon_1 - \varepsilon_3}{\varepsilon_1 + \varepsilon_3} \sum_{n=0}^{\infty} \frac{e^2 \xi^n}{[(\vec{r}_a - \vec{r}_b)^2 + (z_a + z_b - (2n+1)t)^2]^{1/2}} , \tag{8}$$

With random phase approximation, the dielectric polarization function $\varepsilon(q, T)$ can be derived from

$$\varepsilon(q, T) = 1 + \frac{e^2 \Pi(q, T)}{2\varepsilon_0 \varepsilon_1 q} \int_{-t/2}^{t/2} \int_{-t/2}^{t/2} g(z_a, b) g(z_b, b) \, \mathrm{Fr}[F_{CC}(\vec{\rho}_a, \vec{\rho}_b)] dz_a dz_b , \tag{9}$$

where $\varepsilon_0$ is the vacuum permittivity, $q = 2k \sin \frac{\theta}{2}$ is the scattering vector with $k$, $\theta$ being the carrier momentum and the scattering angle, $\Pi(q, T)$ is the 2D finite-temperature electron polarizability; Fr[ ] denotes the 2D Fourier transformation from real space to momentum space. Substituting Eq. 7 into Eq. 9 and merging the terms, one can obtain the expression of the 2D polarization function

$$\varepsilon(q, T) = 1 + \frac{e^2 \Pi(q, T)}{2\varepsilon_0 \varepsilon_1 q} \left[ F_{ee0}(q, t) + \frac{2 \frac{\varepsilon_1 - \varepsilon_2}{\varepsilon_1 + \varepsilon_2} \frac{\varepsilon_1 - \varepsilon_3}{\varepsilon_1 + \varepsilon_3} F_{ee1}(q, t) + \frac{\varepsilon_1 - \varepsilon_3}{\varepsilon_1 + \varepsilon_3} F_{ee2}(q, t) + \frac{\varepsilon_1 - \varepsilon_2}{\varepsilon_1 + \varepsilon_2} F_{ee3}(q, t)}{1 - \frac{\varepsilon_1 - \varepsilon_2}{\varepsilon_1 + \varepsilon_2} \frac{\varepsilon_1 - \varepsilon_3}{\varepsilon_1 + \varepsilon_3} e^{-2qt}} \right] , \tag{10}$$

where $F_{eei}(q, t)$ ($i = 0, 1, 2, 3$) represents the configurative form factors arising from the asymmetric surroundings. The detailed expressions are

$$F_{ee0}(q, t) = \frac{b(8b^2 + 9bq + 3q^2)}{8(b+q)^3},$$

$$F_{ee1}(q, t) = \frac{b^6 e^{-2t(b+q)} [2 - 2e^{t(b-q)} + 2t(b-q) + t^2(b-q)^2][2 - 2e^{t(b+q)} + 2t(b+q) + t^2(b+q)^2]}{4(b^2 - q^2)^3} ,$$

$$F_{ee2}(q, t) = \frac{b^6 e^{-2bt} [2 - 2e^{t(b-q)} + 2t(b-q) + t^2(b-q)^2]^2}{4(b-q)^6},$$

$$F_{ee3}(q, t) = \frac{b^6 e^{-2t(b+q)} [2 - 2e^{t(b+q)} + 2t(b+q) + t^2(b+q)^2]^2}{4(b+q)^6} . \tag{11}$$

When the right dielectric shares the same dielectric constant with the channel ($\varepsilon_3 = \varepsilon_1$), Eq. 10 can be simplified into the calculating expression for a bilayer system, which is exactly the same as the Ando's model for bulk Si FETs.[69]

Then, we derive the scattering matrix elements for CIs located at different positions. For the planar distributed CIs located at the interfaces of bottom (b) dielectric with the position $z_b$ and the density $n_b$, the scattering matrix element is

$$U_b(q,b,t,z_b) = n_b^{1/2} \frac{e^2 b^3 e^{qz_b}}{4\varepsilon_0 \varepsilon_1 q} \frac{F_{b1}(q,b,t) + \frac{\varepsilon_1 - \varepsilon_3}{\varepsilon_1 + \varepsilon_3} F_{b2}(q,b,t)}{(1 - \frac{\varepsilon_1 - \varepsilon_2}{\varepsilon_1 + \varepsilon_2} \frac{\varepsilon_1 - \varepsilon_3}{\varepsilon_1 + \varepsilon_3} e^{-2qt}) \frac{\varepsilon_1 + \varepsilon_2}{2\varepsilon_1}} \qquad z_b \leq -t/2, \qquad (12)$$

where the form factors $F_{bi}(q,b,t)$ (i = 1 or 2) arise from the presence of a point charge and its first-order image, which have the forms

$$F_{b1}(q,b,t) = -\frac{b^3 e^{-t(b+q)} [2 - 2e^{t(b+q)} + 2t(b+q) + t^2(b+q)^2]}{2(b+q)^3},$$

$$F_{b2}(q,b,t) = -\frac{b^3 e^{-t(b+q)} [2 - 2e^{t(b-q)} + 2t(b-q) + t^2(b-q)^2]}{2(b-q)^3}. \qquad (13)$$

Similarly, one can obtain the expression for the CIs located at the interfaces of the top (t) dielectric

$$U_t(q,b,t,z_t) = n_t^{1/2} \frac{e^2 b^3 e^{qz_t}}{4\varepsilon_0 \varepsilon_1 q} \frac{F_{t1}(q,b,t) + \frac{\varepsilon_1 - \varepsilon_2}{\varepsilon_1 + \varepsilon_2} F_{t2}(q,b,t)}{(1 - \frac{\varepsilon_1 - \varepsilon_2}{\varepsilon_1 + \varepsilon_2} \frac{\varepsilon_1 - \varepsilon_3}{\varepsilon_1 + \varepsilon_3} e^{-2qt}) \frac{\varepsilon_1 + \varepsilon_3}{2\varepsilon_1}} \qquad z_t \geq t/2, \qquad (14)$$

with its form factors

$$F_{t1}(q,b,t) = -\frac{b^3 e^{-bt} [2 - 2e^{t(b-q)} + 2t(b-q) + t^2(b-q)^2]}{2(b-q)^3},$$

$$F_{t2}(q,b,t) = -\frac{b^3 e^{-t(b+2q)} [2 - 2e^{t(b+q)} + 2t(b+q) + t^2(b+q)^2]}{2(b+q)^3}. \qquad (15)$$

The different forms of the matrix elements $U_b$ and $U_t$ are a consequence of the lopsided carrier distribution.

According to the Boltzmann theory, the rate of elastic scattering in 2D systems is given by

$$\frac{1}{\tau_j(k)} = \frac{2D_0}{\hbar g_s g_v} \int_0^\pi \left| \frac{U_j(q)}{\varepsilon(q,T)} \right|^2 (1 - \cos\theta) \mathrm{d}\theta \qquad , \qquad (16)$$

where the index j denotes different elastic scattering centers, $D_0 = \frac{g_s g_v m^*}{2\pi\hbar^2}$ is the 2D density of states with $\hbar$ and $m^*$ being the reduced Planck constant and effective mass, and $g_s$ and $g_v$ are the spin and valley degeneracy factors, respectively. All the configurative details are reflected in the form factors $F_{vb}(q,b,t)$ and passed to the scattering matrix elements $U_j(q)$ and the dielectric polarization function $\varepsilon(q,T)$. Note that there are controversies concerning the values of $m^*$ and $g_v$ for few-layer MoS$_2$ flakes.[74,75] For simplicity, we assume constant $D_0$ and $g_v$ in the calculations, which also allows us to solely study the effect of channel thickness on the intensity of impurity scattering.

Apart from Coulomb impurity, lattice phonons are also important scattering mechanisms at room temperature. In compound semiconductors, two types of phonon scattering mechanisms exist, including the deformation potential and the Fröhlich interaction. On the one hand, lattice vibration can perturb the periodic lattice potential and scatter off the electron waves through the deformation potentials, which is known as the deformation potential scattering. To discern the phonon attributes, the scattering processes can be further divided into acoustic deformation phonon (ADP) and optical deformation phonon (ODP) according to their vibration modes. On the other hand, the vibration of optical polar phonons gives rise to a macroscopic electric field that can couple to the charge carriers, resulting in the Fröhlich interaction.

To compare the scattering intensities of different sources, lattice phonons are also calculated.[76-79] Following the calculation of Kaasbjerg $et\ al.$[77], the carrier screening effect is not included in the ADP and ODP mechanisms, and thus they are independent of thickness. In contrast, we consider the carrier screening effect in the Fröhlich interaction and CI scattering such that they are functions of channel thickness $t$. By incorporating configurative form factors into the $t$-dependent coefficients $\alpha_b(t)$ and $\alpha_t(t)$, the total scattering rate ($\tau$) can then be written as

$$\tau(t)^{-1} = \alpha_b(t)n_b + \alpha_t(t)n_t + \beta_F(t) + \beta_{ADP} + \beta_{ODP}, \qquad (17)$$

where $n_b$ and $n_t$ are the interfacial impurity densities at the bottom and top channel surfaces, respectively, and $\beta_F(t)$ is the phonon contribution from the Fröhlich interaction, which is moderately $t$-dependent. $\beta_{ADP}$ and $\beta_{ODP}$ denote the contribution from the $t$-independent ADP and ODP, respectively. The coefficients for two device gating structures are plotted in Fig. 7, with the back-gated air/MoS$_2$/SiO$_2$ FET structure in Fig. 7(a) and the top-gated HfO$_2$/MoS$_2$/SiO$_2$ FET structure in Fig. (b). In both cases, the scattering rates of CIs show a more pronounced dependence on thickness than those of phonons. When the CI densities are high, a strong thickness-dependent device performance is naturally observed.

The underlying reason why we can use CI scattering to interpret the strong mobility degradation in monolayers is that the scattering rate $\tau$ increases with the shrunk interaction distance $d$, because $\tau_{CI} \propto V(d)^2 \propto d^{-2}$ roughly, where $V(d)$ represents the Coulomb potential. As an instance, Figure 8(b) schematically illustrates and compares the interaction distances of the top and bottom ($d_{tN}$ and $d_{bN}$, with N the number of MoS$_2$ layers) channel surfaces between 1- and 5-layer channels. The red dots and circular shades denote the interfacial charged impurities and corresponding scattering potential, respectively. As the channel thickness decreases from 5 to 1 layer, both the $d_t$ and $d_b$ are reduced and hence the carrier scattering is intensified. Figure 8(c) shows a plot of individual $\tau$ components for phonons[76-78] and CIs for a back-gated MoS$_2$ FET by assuming $n_t = n_b = 3 \times 10^{12} \text{cm}^{-2}$. The CI contribution from the gated (bottom) surface dominates in the thickness range from 1 to 10 layers. In contrast, the contribution of the ungated (top) surface is strong only for thin channels ($\leq 3$ layers) and becomes weak or even negligible for thick channels. The distinct dependences of CIs located at the top and bottom channel surfaces stem from different variation trends of $d_b$ and $d_t$ along with channel thickness, where $d_b$ only slightly increases with thickness while $d_t$ changes more markedly. In Fig. 8(d), we compare the theoretical calculation with the experiment. The experimental mobility data correspond to $n_b \sim 3 \times 10^{12} \text{cm}^{-2}$, indicating that typical samples contain a high density of CIs, owing to the gaseous absorbates on the channel surfaces, and the trapped charges and chemical bonds on the SiO2 dielectrics. Evidently, the reduction in the interaction distance between interfacial CIs and channel carriers along with channel thickness is responsible for the heavy thickness dependence and low performance of ultrathin devices.

## 3. Conclusions and Outlook

We performed experimental and theoretical studies of the origins of the dependence of carrier mobility on thickness. We revealed that the expanded injection barrier at contacts with decreasing thickness and the interfacial Coulomb impurities are the main factors responsible for the observed thickness dependence.

With the above results in mind, several technological suggestions are proposed. First, creating more electrically transparent contacts is necessary for improving device performance. In this regard, the energy-level matching between the metallic electrodes and the channels is particularly important.[80-82] Contact engineering with degenerate doping would further lower the barrier width and result in efficient charge injection.[28,29] Besides, semiconductors with a low bulk bandgap should be considered as channel candidates, considering the bandgap expansion after thinning down. Hence, the search for different channel materials with a technologically suitable bandgap is also necessary. Second, developing an encapsulation technique for ultrathin channels in order to reduce CI density at channel surfaces is also critical. For instance, isolating the channel using clean encapsulators such as BN,[80-82] PMMA,[27,30] and SAM layers[33] has proved effective for improving the device performance considerably. With appropriate optimization, it is expected that technologically viable atomic FETs based on 2D semiconductors will be developed for next-generation nanoelectronics.

## Acknowledgements

The authors thank Dr. Katsunori Wakabayashi, Dr. Shu Nakaharai, Dr. Yong Xu, Dr. Katsuyoshi Komatsu, and Dr. Mahito Yamamoto for their fruitful discussion and comments. This research was supported by a Grant-in-Aid (Kakenhi Nos. 25107004) from the Japan Society for the Promotion of Science (JSPS).

# Figure 1

**a**

S⁻²
Mo⁺⁴

$V_d$

Drain
SiO₂
Source

Back gate (Si)
$V_g$

**b**

FET, Ref. 61
FET, Ref. 60
Hall, Ref. 59

Hall

FET

Carrier mobility (cm²V⁻¹s⁻¹)

Number of MoS₂ layers

Bulk

Figure 2

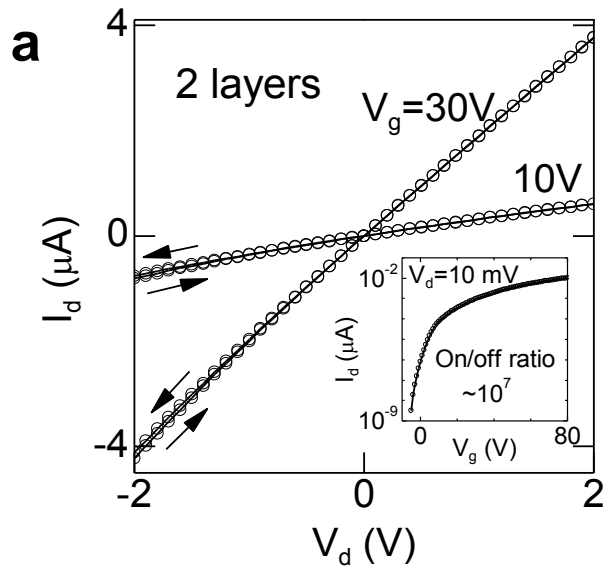

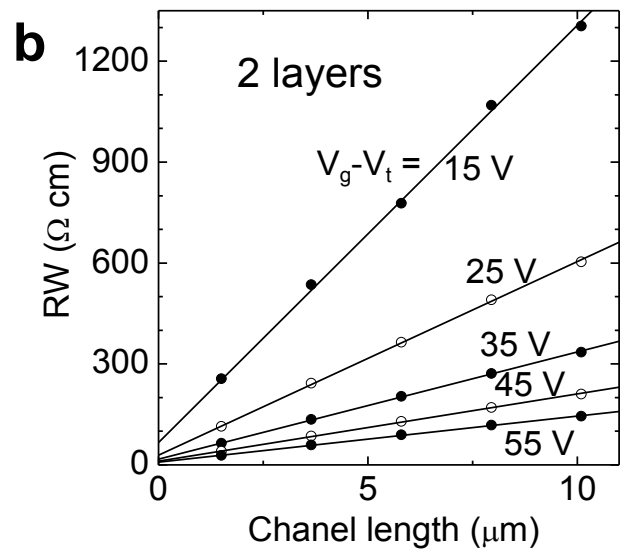

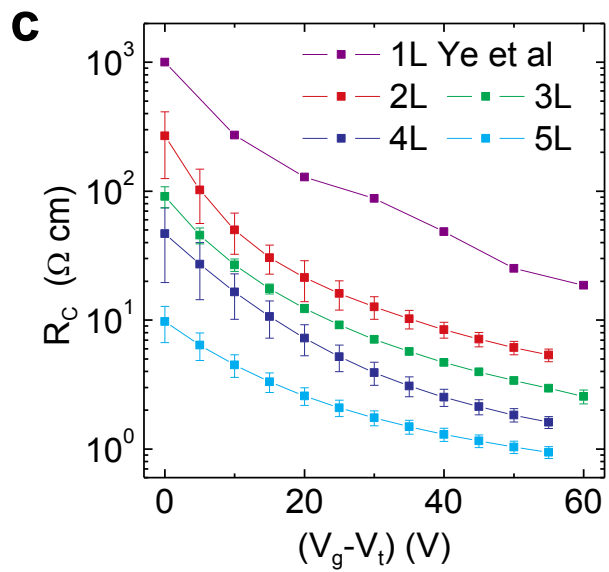

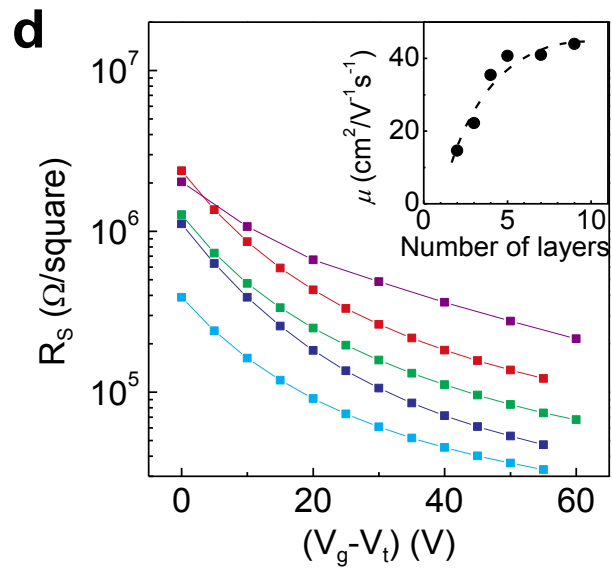

Figure 3

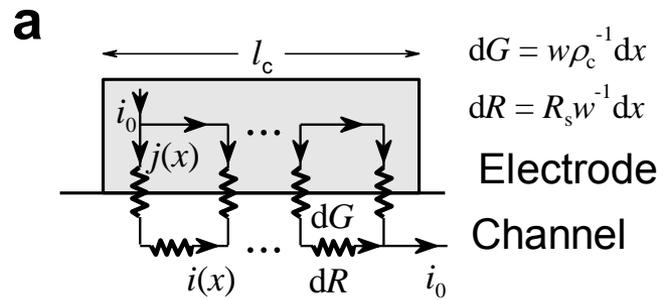

$\mathrm{d}G = w\rho_\mathrm{c}^{-1}\mathrm{d}x$

$\mathrm{d}R = R_\mathrm{s}w^{-1}\mathrm{d}x$

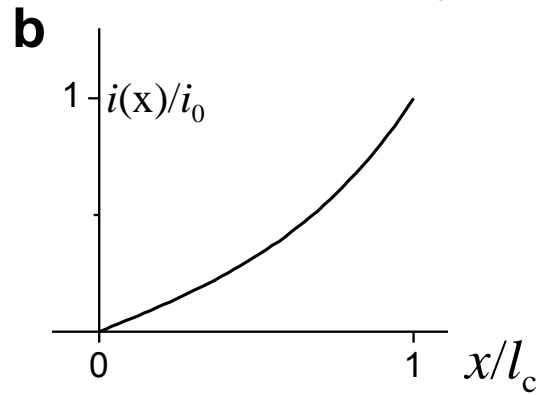

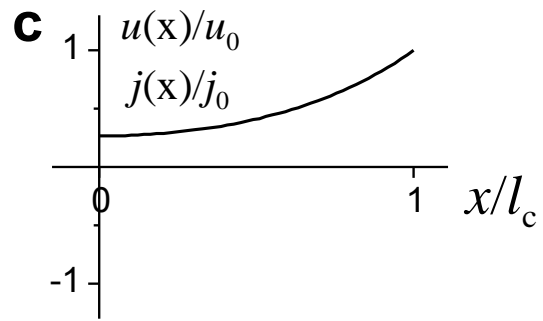

Figure 4

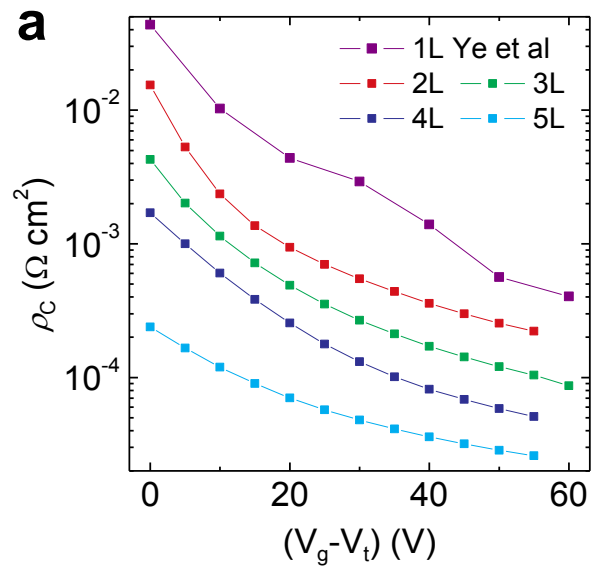

**a**

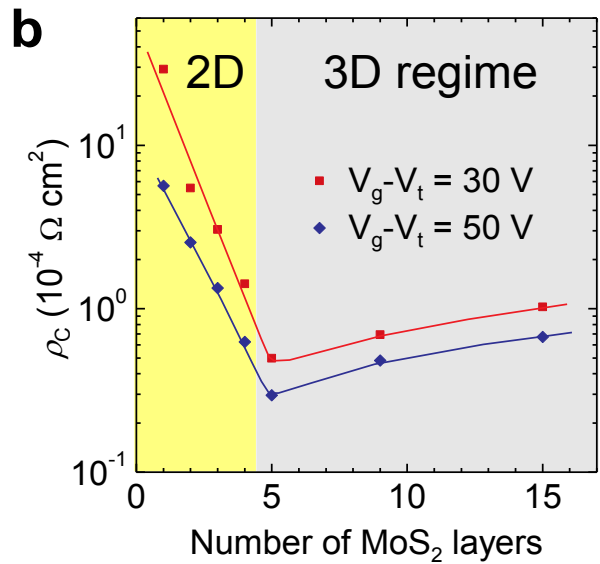

**b**

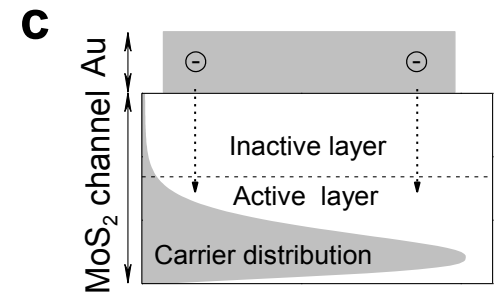

**c**

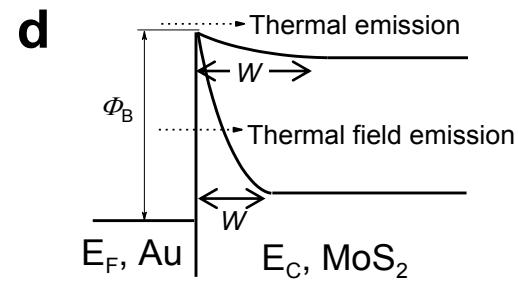

**d**

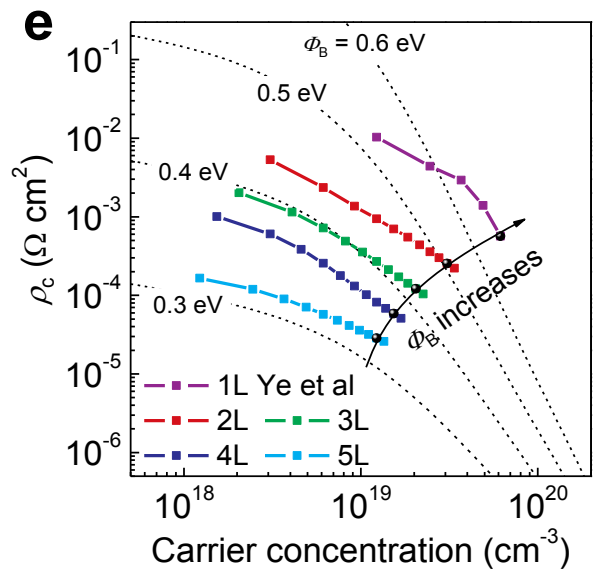

**e**

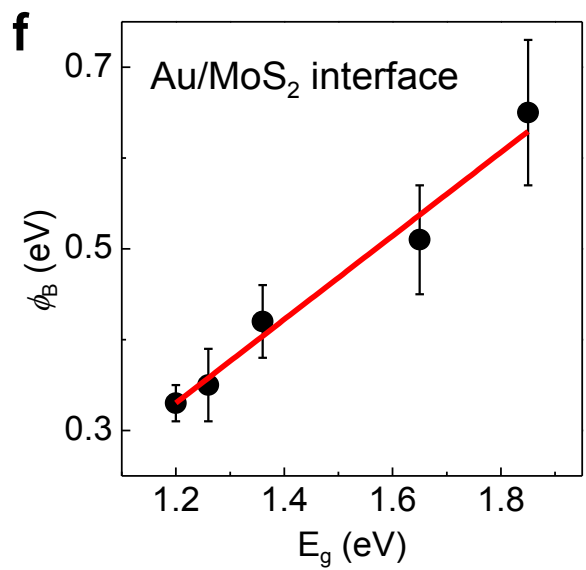

**f**

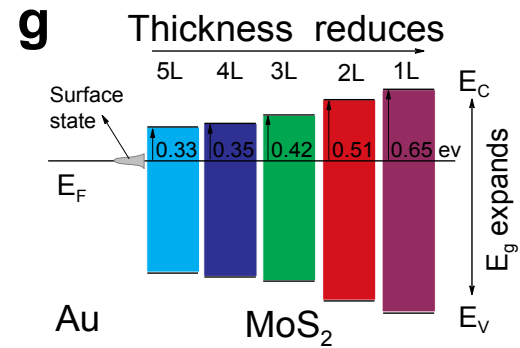

**g**

Figure 5

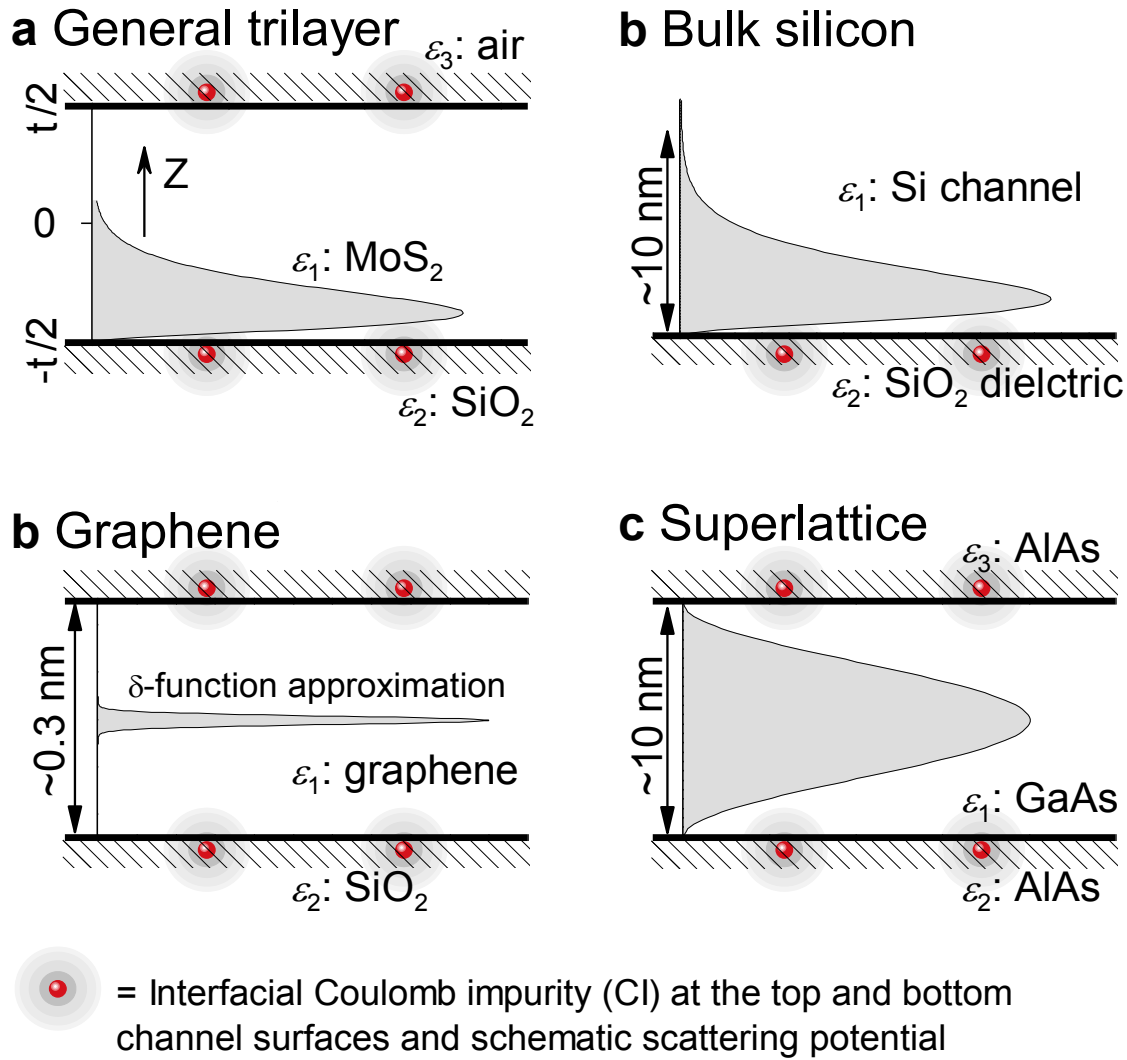

**a** General trilayer $\varepsilon_3$: air

t/2

0 Z

-t/2

$\varepsilon_1$: MoS$_2$

$\varepsilon_2$: SiO$_2$

**b** Bulk silicon

~10 nm $\varepsilon_1$: Si channel

$\varepsilon_2$: SiO$_2$ dielctric

**b** Graphene

~0.3 nm $\delta$-function approximation

$\varepsilon_1$: graphene

$\varepsilon_2$: SiO$_2$

**c** Superlattice $\varepsilon_3$: AlAs

~10 nm

$\varepsilon_1$: GaAs

$\varepsilon_2$: AlAs

= Interfacial Coulomb impurity (CI) at the top and bottom channel surfaces and schematic scattering potential

Figure 6

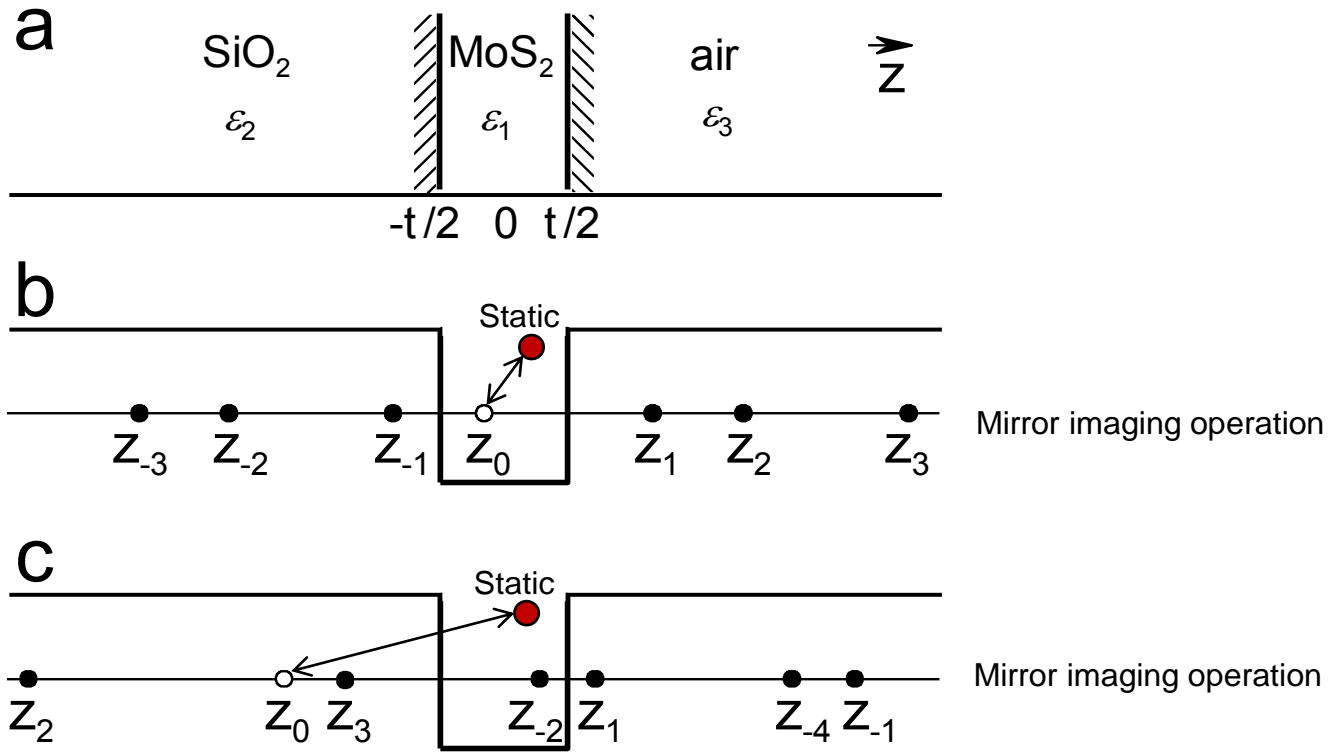

Mirror imaging operation

Mirror imaging operation

Figure 7

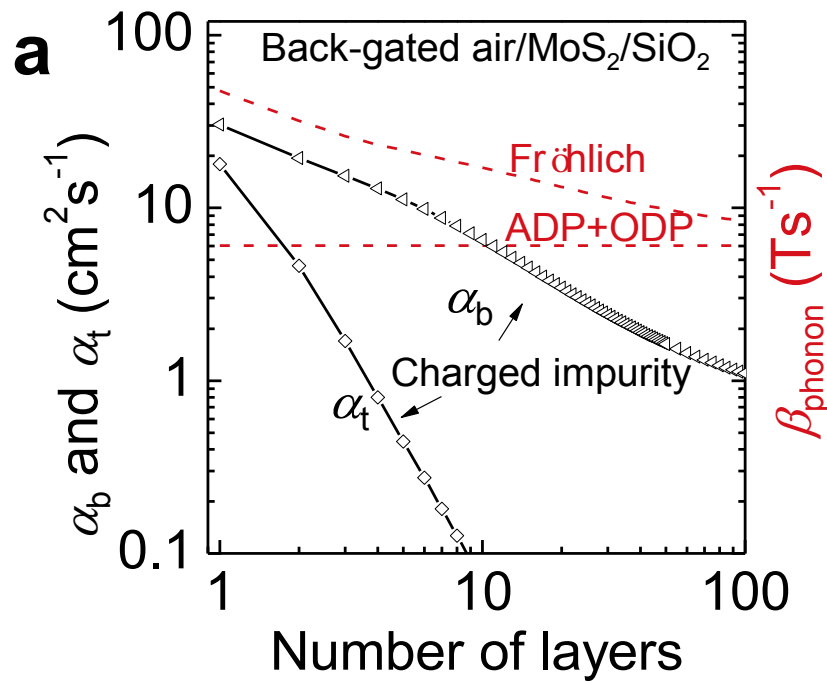
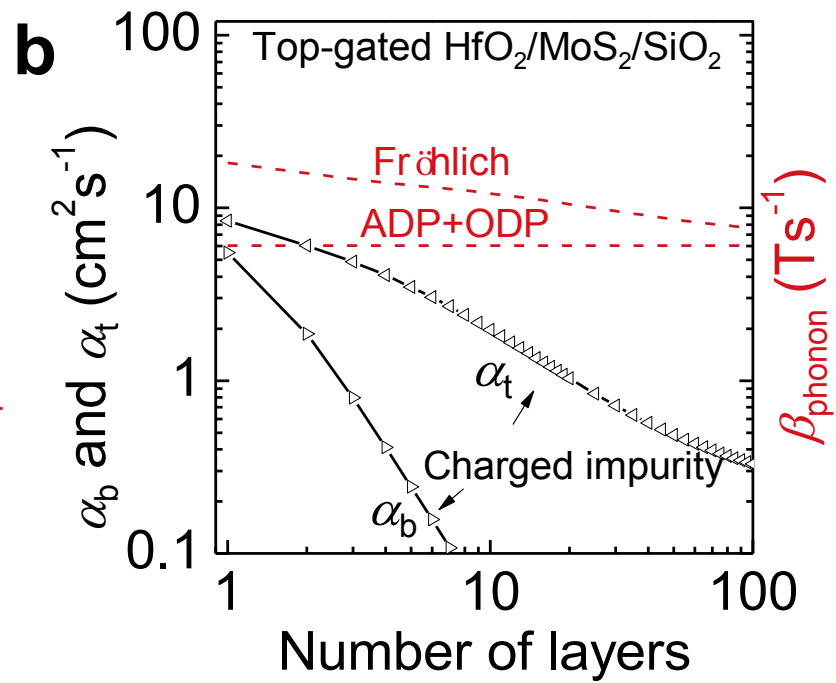

Figure 8

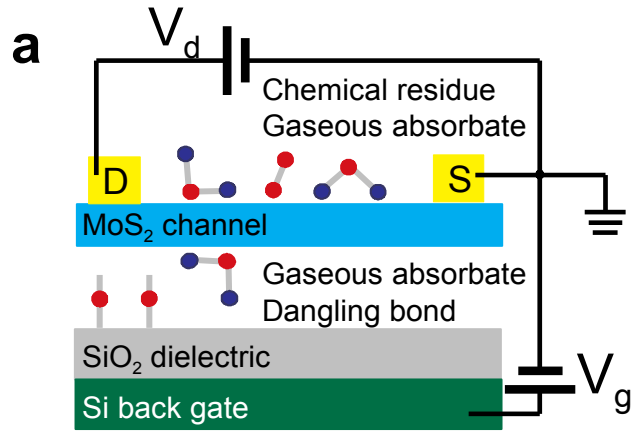

**a**

$V_d$

Chemical residue
Gaseous absorbate

D    S

MoS₂ channel

Gaseous absorbate
Dangling bond

SiO₂ dielectric

Si back gate

$V_g$

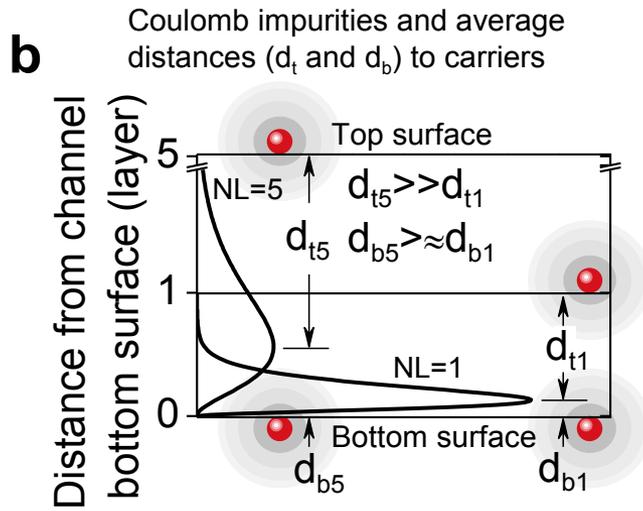

**b**    Coulomb impurities and average
distances ($d_t$ and $d_b$) to carriers

Distance from channel bottom surface (layer)

Top surface

$d_{t5} \gg d_{t1}$

NL=5

$d_{t5}$    $d_{b5} > \approx d_{b1}$

NL=1

$d_{t1}$

Bottom surface

$d_{b5}$    $d_{b1}$

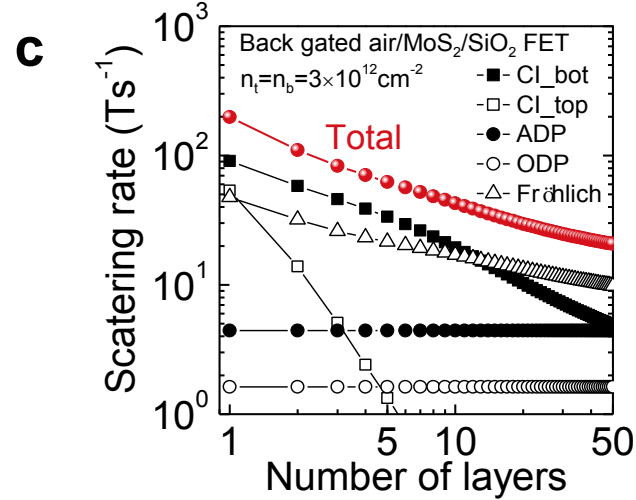

**c**    Back gated air/MoS₂/SiO₂ FET
$n_t = n_b = 3 \times 10^{12} cm^{-2}$

Scatering rate (Ts⁻¹)

■ Cl_bot
□ Cl_top
● ADP
○ ODP
△ Fröhlich

Total

Number of layers

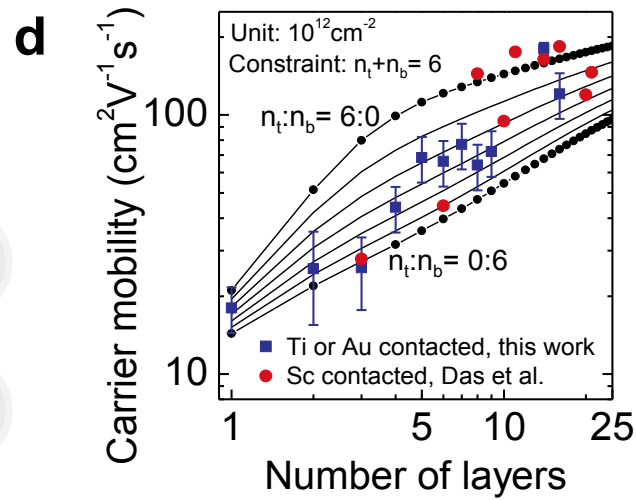

**d**    Unit: $10^{12} cm^{-2}$
Constraint: $n_t + n_b = 6$

Carrier mobility (cm²V⁻¹s⁻¹)

$n_t : n_b = 6:0$

$n_t : n_b = 0:6$

■ Ti or Au contacted, this work
● Sc contacted, Das et al.

Number of layers

**Figure captions**

**Fig. 1.** (a) Schematic diagram and optical image of an FET device with electrode layout for transfer line measurement. The inset shows the atomic structure for bilayer $MoS_2$. Adapted from Ref. 63. Copyright 2014 American Chemical Society. (b) Carrier mobility ($\mu$) as a function of channel thickness. The dashed lines are guides for the eyes. Adapted from Ref. 61. Copyright 2013 American Chemical Society.

**Fig. 2.** (a) Typical electrical properties of bilayer $MoS_2$ FETs. The inset is the corresponding transfer curve showing an on/off current ratio of $10^7$. (b) Transfer line plot for extracting contact resistivity ($R_c$) and sheet resistivity ($R_s$) under different gating conditions. (c) and (d) Extracted $R_c$ and $R_s$ values for different sample thicknesses in unit of number of $MoS_2$ layers (NL with N being an integer). Inset in (d): Intrinsic carrier mobility ($\mu$) *vs* channel thickness. Adapted from Ref. 63. Copyright 2014 American Chemical Society.

**Fig. 3.** Resistor network model for analyzing current distribution at electrode/channel contact area. (a) Impedance elements in the electrode/channel stack. (b) and (c) Schematic illustration of lateral channel current $i(x)$, vertical potential $u(x)$, and injection density $j(x)$. Adapted from Ref. 63. Copyright 2014 American Chemical Society.

**Fig. 4.** (a) and (b) Gate and thickness dependences of specific contact resistivity ($\rho_c$). (c) Schematic carrier distribution and injection path for a back-gated FET with thick channel (*i.e.*, 3D transport regime). (d) Schematic diagram of band alignments and carrier injection paths for the thermal emission (TE) and thermal field emission (TFE) injection theories at the electrode/channel contacts. The difference between them lies in the width of the interfacial barrier, which changes with the gate bias and the semiconductor carrier density. (e) Comparison of $\rho_c$ data (dotted lines) with theoretical results of TFE injection mechanism (dashed lines) to extract barrier heights. (f) Thickness scaling effect on the barrier height ($\phi_B$) at $Au/MoS_2$ contacts, which is a function of semiconductor bandgap ($E_g$). (g) Evolution of energy level alignment around $Au/MoS_2$ contacts as channel thickness decreases. Adapted from Ref. 63. Copyright 2014 American Chemical Society.

**Fig. 5.** Schematic diagrams of dielectric surroundings and carrier distributions for different device configurations. (a) A common trilayer structure: (1) two boundaries that produce infinite mirror imaging charges; (2) a lopsided carrier distribution that leads to complicated configurative form factors in scattering matrix elements and dielectric polarization function. (b) Bulk silicon: one boundary that produces only one mirror imaging charge.[69] (c) Graphene: negligible channel thickness $t$ for the middle layer and a simple pulselike carrier distribution with a Dirac $\delta$ function.[72,73] (d) Superlattice: symmetric dielectrics and trigonometric wavefunction.[70,71] Adapted from Ref. 61. Copyright 2013 American Chemical Society.

**Fig. 6.** (a) Dielectric surroundings for a common trilayer structure, where the semiconductor channel is sandwiched by two asymmetric dielectrics. (b) and (c) Applying the mirror imaging method to derive the Coulomb force in the trilayer structure where the two boundaries produce infinite mirror imaging charges. In (b), two point charges are located in the channel. One charge is fixed and the other charge is mirrored through the two dielectric boundaries. The positions of the mirror imaging charges are $z_n = nt + (-1)^n z_0$, n=0, $\pm 1, \pm 2$ ... In (c), one charge is in the channel and the other is in the left dielectric. Adapted from Ref. 61. Copyright 2013 American Chemical Society.

**Fig. 7.** Values of Coulomb impurity (CI) and phonon scattering coefficients $\alpha_b$, $\alpha_t$, and $\beta_{phonon}$ for (a) back-gated air/MoS$_2$/SiO$_2$ and (b) top-gated HfO$_2$/MoS$_2$/SiO$_2$ FETs. Adapted from Ref. 61. Copyright 2013 American Chemical Society.

**Fig. 8.** (a) Diagram of charged impurities (*e.g.*, chemical residues, gaseous adsorbates, and surface dangling bonds) located on the top and bottom channel surfaces, which are the leading scatterers in ultrathin channels. (b) Comparison of interaction distances $d_t$ and $d_b$ between 1- and 5-layer channels. The red dots and circular shades denote the interfacial charged impurities and corresponding scattering potential, respectively. (c) Calculated scattering rates for a back-gated air/MoS$_2$/SiO$_2$ structure by assuming $n_t = n_b = 3 \times 10^{12}\,cm^{-2}$. (d) Comparison of calculation and experiment. Adapted from Ref. 61. Copyright 2013 American Chemical Society.